\documentclass[oneside,english]{autart}
\usepackage[T1]{fontenc}
\usepackage[latin9]{inputenc}
\usepackage{color}
\usepackage{amstext}
\usepackage{amssymb}
\usepackage{graphicx}

\makeatletter
\makeatletter
\def\old@comma{,}
\catcode`\,=13
\def,{%
  		\ifmmode%
    		\old@comma\discretionary{}{}{}%
  		\else%
    		\old@comma%
  		\fi%
}
\makeatother
\usepackage{cite}
\usepackage{url}
\usepackage{amssymb}

\makeatother

\usepackage{babel}
\begin{document}
\begin{frontmatter}

\title{Containment Control for a Social Network with State-Dependent Connectivity}

\author{}\author[ufl]{Zhen Kan}\ead{kanzhen0322@ufl.edu},
\author[ufl]{Justin Klotz}\ead{jklotz@ufl.edu},
\author[reef]{Eduardo L. Pasiliao Jr}\ead{pasiliao@eglin.af.mil},
\author[ufl]{Warren E. Dixon}\ead{wdixon@ufl.edu}
\address[ufl]{Department of Mechanical and Aerospace Engineering, University of Florida, Gainesville, USA}
\address[reef]{Air Force Research Laboratory, Munitions Directorate, Eglin AFB, FL 32542, USA.}
\thanks{This research is supported in part by NSF award numbers 0547448, 0901491, 1161260, and a contract with the Air Force Research Laboratory, Munitions Directorate at Eglin AFB. }
\begin{abstract}
Social interactions influence our thoughts, opinions and actions.
In this paper, social interactions are studied within a group of individuals
composed of influential social leaders and followers. Each person
is assumed to maintain a social state, which can be an emotional state
or an opinion. Followers update their social states based on the states
of local neighbors, while social leaders maintain a constant desired
state. Social interactions are modeled as a general directed graph
where each directed edge represents an influence from one person to
another. Motivated by the non-local property of fractional-order systems,
the social response of individuals in the network are modeled by fractional-order
dynamics whose states depend on influences from local neighbors and
past experiences. A decentralized influence method is then developed
to maintain existing social influence between individuals (i.e., without
isolating peers in the group) and to influence the social group to
a common desired state (i.e., within a convex hull spanned by social
leaders). Mittag-Leffler stability methods are used to prove asymptotic
stability of the networked fractional-order system.
\end{abstract}
\end{frontmatter}

\section{Introduction}

Social interactions influence our emotions, opinions, and behaviors.
Technological advances in social media provide more rapid, convenient,
and widespread communication among individuals, which leads to a more
dynamic interaction and influence. For example, recent riots \cite{Bright2011}
and ultimately revolution \cite{Gustin2011}, have been facilitated
through social media technologies such as Facebook, Twitter, YouTube,
and BlackBerry Messaging (BBM). Marketing agencies also have begun
to take advantage of influence due to social media, especially through
the internet. The company Razorfish, for example, works with online
peer influencers to transform them into brand advocates through the
execution of Social Influence Marketing (SIM) Strategy, which aims
to influence marketing primarily through online, small groups, peer
pressure, reciprocity or flattery \cite{Singh2009}.

Various dynamic models have been developed to study the individual's
social behavior, such as the efforts to model the emotional response
of different individuals \cite{sprott2004love,sprott2005happiness,ghosh2010fear}.
In \cite{sprott2004love}, the time-variation of emotions between
individuals involved in a romantic relationship is described by a
dynamic model of love, and in \cite{sprott2005happiness} a set of
differential equations are developed to model the individual's happiness
in response to exogenous influences. Fractional-order differential
equations are a generalization of integer-order differential equations,
and they exhibit a non-local integration property where the next state
of a system not only depends upon its current state but also upon
its historical states starting from the initial time \cite{monje2010}.
Motivated by this property, many researchers have explored the use
of fractional-order systems as a model for various phenomena in natural
and engineered systems. For instance, the works in \cite{sprott2004love}
and \cite{sprott2005happiness} were revisited in \cite{Ahmad2007}
and \cite{song2010happiness}, where the models of love and happiness
were generalized to fractional-order dynamics by taking into account
the fact that a person's emotional response is influenced by past
experiences and memories. However, the models developed in \cite{sprott2004love,sprott2005happiness,Ahmad2007,song2010happiness}
only focus on an individual's emotional response, without considering
the interaction with social peers where rapid and widespread influences
from social peers can prevail. \textcolor{black}{Other results, such
as \cite{Cucker2007,Blondel2009} and the reference therein, studied
the interaction of social peers using an opinion dynamics model, and
derived conditions under which consensus can be reached. However,
agents in \cite{Cucker2007,Blondel2009} only update their opinions
by averaging the neighboring agent opinions, without taking into account
the influence of agents' past experience and memory on their decision
making.}

When making a decision or forming an opinion, individuals tend to
communicate with parents, friends, or colleagues and take advice from
social peers. Social connections such as friendship, kinship, and
other relationships can influence the decisions they make. Some individuals
(e.g., parents, teachers, mentors, and celebrities) may exhibit more
powerful influences in others' decision making, and the underlying
social network enables the influence to pass from influential individuals
to receptive individuals. Containment control is a particular class
of consensus problems (see \cite{Ren2007a,Olfati-Saber2007} for a
comprehensive literature review for consensus problems), in which
follower agents are under the influence of leaders through local information
exchange in a leader-follower network. In results such as \cite{Notarstefano2011,Cao2009,Mei2012,Cao2012},
distributed containment control algorithms are developed for agents
with integer-order dynamics where the group of followers is driven
to a convex hull spanned by multiple leaders' states under an undirected,
directed or switching topology. This paper examines how such methods
can be leveraged to manipulate a social network. This work specifically
aims to investigate how peer pressure from social leaders affects
consensus beliefs (e.g., opinions, emotional states, purchasing decisions,
political affiliation, etc.) within a social network, and how an interaction
algorithm can be developed such that the group social behavior can
be driven to a desired end (i.e., a convex hull spanned by the leaders'
states).

By modeling human emotional response as a fractional-order system,
the influence of a person's emotions within a social network is studied,
and emotion synchronization for a group of individuals is achieved
in our recent preliminary work \cite{Kan.Shea.ea2012,Kan.Klotz.ea2013}.
However, the emotion synchronization behavior in \cite{Kan.Shea.ea2012}
only considers an undirected network structure: the one-sided influence
of social leaders is not considered. This work aims to investigate
how the social beliefs (e.g., emotional response, opinions, etc.)
of a group of individuals evolve under the influence of social leaders.
Similar to \cite{Kan.Shea.ea2012}, the social group is modeled as
a networked fractional-order system, where the social response of
each individual is described by fractional-order dynamics whose states
depend on influences from social peers, as well as past experiences.
Since social leaders are considered, the undirected network topology
in \cite{Kan.Shea.ea2012} is extended to a directed graph, where
the directed edges indicate the influence capability between two individuals
(e.g., the leaders can influence the followers' state, but not vice
versa). The goal in this work is to develop a decentralized influence
algorithm where individuals within a social group update their beliefs
by considering beliefs from social peers and the social group achieves
a desired common belief (i.e., the social state of the group converges
to a convex hull spanned by social leaders). Since an individual generally
only considers others' beliefs as reasonable if their beliefs differ
by less than a threshold, social difference is introduced to measure
the closeness of the beliefs between individuals. In contrast to the
constant weights considered in \cite{Notarstefano2011,Cao2009,Mei2012},
the social difference is a time-varying weight which depends on individuals'
current states. Moreover, instead of assuming network connectivity
(i.e., there always exists a path of influence between any two agents)
such as in \cite{Notarstefano2011,Cao2009,Mei2012}, one main challenge
here is to influence the followers' social states to a desired end
by maintaining consistent interaction among social peers and influential
leaders (i.e., individuals can always be influenced by social peers,
instead of being isolated from the social group) within a time-varying
graph. When modeled as a networked fractional-order system, the development
of a containment algorithm can be more challenging compared to the
integer-order dynamics in \cite{Notarstefano2011,Cao2009,Mei2012,Cao2012},
which can be considered as a particular case of generalized fractional-order
dynamics. The first apparent result that investigated the coordination
of networked fractional systems is \cite{cao2010Fractional}. However,
only linear time-invariant systems are considered in \cite{cao2010Fractional},
where the interaction between agents is modeled as a link with a constant
weight. Due to the time-varying weights considered here, previous
stability analysis tools such as examining the eigenvalues of linear
time-invariant fractional-order systems (cf. \cite{cao2010Fractional,chen2006robust,song2010happiness})
are not applicable to the time-varying networked fractional-order
system in this work. To address these challenges, a decentralized
influence function is developed to achieve containment control for
the networked fractional-order systems while preserving continued
social interaction among individuals. Asymptotic convergence of the
social states to the convex hull spanned by leaders' states in the
social network is then analyzed via LaSalles's invariance theorem
\cite{Khalil2002}, convex properties \cite{Boyd2004} and a Mittag-Leffler
stability \cite{li2009mittag} approach.

\section{Preliminaries\label{sec:preliminary}}

Consider a Fractional Order System (FOS)
\begin{equation}
_{t_{0}}\mathcal{D}_{t}^{\alpha}x\left(t\right)=f\left(t,x\right)\label{fractional}
\end{equation}
with initial condition%
\footnote{\textcolor{black}{The initial condition $x\left(t_{0}\right)$ is
defined as a linear combination of internal states $z_{k}\left(t_{0}\right)$,
$k=1,\ldots,J,$ where $z_{k}\left(t_{0}\right)$ contains all historical
information of the system for $t<t_{0}$ based on the work in \cite{Trigeassou2011}.
The infinite state model approach to resolve the initialization in
\cite{Trigeassou2011} is also used in the subsequent simulation section.}%
} $x\left(t_{0}\right)$, where $_{t_{0}}\mathcal{D}_{t}^{\alpha}$
denotes the fractional derivative operator with order $\alpha\in(0,1]$
on a time interval $[t_{0},t]$, and $f\left(t,x\right)$ is piecewise
continuous in $t$ and locally Lipschitz in $x$. Similar to the exponential
function used in solutions of integer-order differential equations,
the Mittag-Leffler (M-L) function given by $E_{\alpha,\beta}\left(z\right)=\sum{}_{k=0}^{\infty}\frac{z^{k}}{\Gamma\left(k\alpha+\beta\right)}$,
where $\alpha,$ $\beta>0$ and $z\in$ $\mathbb{C}$, is frequently
used in solutions of fractional-order systems. Particularly, when
$\alpha=\beta=1,$ $E_{1,1}\left(z\right)=e^{x}$ is an exponential
function$.$ Stability of the solutions to (\ref{fractional}) are
defined by the M-L function as follows \cite{li2009mittag}.
\begin{defn}
\label{def1}\cite{li2009mittag} (Mittag-Leffler Stability) The solution
of (\ref{fractional}) is said to be Mittag-Leffler stable if $\left\Vert x\left(t\right)\right\Vert \leq\left\{ m\left[x\left(t_{0}\right)\right]E_{\alpha,1}\left(-\lambda\left(t-t_{0}\right)^{\alpha}\right)\right\} ^{b},$
where $t_{0}$ is the initial time, $\alpha\in\left(0,1\right),$
$b>0,$ $\lambda>0,$ $m\left(0\right)=0,$ $m\left(x\right)\geq0,$
$m\left(x\right)$ is locally Lipschitz, and $E_{\alpha,1}$ is defined
in $E_{\alpha,\beta}$ with $\beta=1.$ 
\end{defn}
Lyapunov's direct method is extended to fractional-order systems in
the following Lemma to determine Mittag-Leffler stability for the
solutions of (\ref{fractional}) in \cite{li2009mittag}.
\begin{lem}
\label{lemma2}\cite{li2009mittag} Let $x=0$ be an equilibrium point
for the fractional order system (\ref{fractional}), and $\mathbb{D\subset R}^{n}$
be a domain containing the origin. Let $V\left(t,x\right):\left(0,\infty\right]\times\mathbb{D\rightarrow R}$
be a continuously differentiable function and locally Lipschitz with
respect to $x$ such that
\begin{eqnarray*}
\sigma_{1}\left(\left\Vert x\right\Vert \right) & \leq & V\left(t,x\right)\leq\sigma_{2}\left(\left\Vert x\right\Vert \right),\\
_{0}D_{t}^{\beta}V\left(t,x\right) & \leq & -\sigma_{3}\left(\left\Vert x\right\Vert \right),
\end{eqnarray*}
where $x\in\mathbb{D}$, $\beta\in\left(0,1\right),$ and $\sigma_{i}$
($i=1,2,3$) are class $\mathcal{K}$ functions%
\footnote{A continuous function $\sigma:\left[0,a\right)\rightarrow\left[0,\infty\right)$
is said to belong to class $\mathcal{K}$ if it is strictly increasing
and $\sigma\left(0\right)=0.$ It is said to belong to class $\mathcal{K}_{\infty}$
if $a=\infty$ and $\sigma\left(r\right)\rightarrow\infty$ as $r\rightarrow\infty$
\cite{Khalil2002}.%
}. Then $x=0$ is Mittag-Leffler stable, which implies that the equilibrium
point of (\ref{fractional}) is asymptotically stable. 
\end{lem}

\begin{defn}
\label{def3}\cite{Boyd2004} For a set of points $X\triangleq\left\{ x_{1},\cdots,x_{n}\right\} ,$
the convex hull $Co\left(X\right)$ is defined as the minimal set
containing all points in $X$, satisfying that $Co\left(X\right)\triangleq\left\{ \sum\nolimits _{i=1}^{n}\alpha_{i}x_{i}\left\vert x_{i}\in X,\text{ }\alpha_{i}>0,\mbox{ }\sum\nolimits _{i=1}^{n}\alpha_{i}=1\right.\right\} $. 
\end{defn}
Graph theory (cf. \cite{RM_LAA:94} and \cite{Mesbahi2010}) is widely
used to represent a networked system. Let $\mathcal{G}=(\mathcal{V},\mathcal{E})$
denote a directed graph, where $\mathcal{V=}\left\{ v_{1},\cdots,v_{N}\right\} $
and $\mathcal{E}$ $\subset\mathcal{V\times V}$ denote the set of
nodes and the set of edges, respectively. Each edge $\left(v_{i},v_{j}\right)\in\mathcal{E}$
represents the neighborhood of node $i$ and node $j$, which indicates
that node $i$ is able to access states of node $j$, but not vice
versa. The neighbor set of node $i$ is denoted as $\mathcal{N}_{i}=\left\{ v_{j}\text{ }|\text{ }\left(v_{i},v_{j}\right)\in\mathcal{E}\right\} .$
A directed path from node $v_{1}$ to node $v_{k}$ is a sequence
of edges $\left(v_{1},v_{2}\right),$ $\left(v_{2},v_{3}\right),\cdots,$
$\left(v_{i},v_{k}\right)$ in the directed graph $\mathcal{G}$.
If graph $\mathcal{G}$ contains a directed tree, every node has exactly
one parent node except for one node, called the root, and the root
has directed paths to every other node in graph $\mathcal{G}$. The
adjacency matrix is defined as $A\triangleq\left[a_{ij}\right]\in\mathbb{R}^{N\times N}$
with $a_{ij}>0$ if $\left(v_{i},v_{j}\right)\in\mathcal{E}$, and
$a_{ij}=0$ otherwise, where $a_{ij}$ represents a weighting factor
for the associated edge $\left(v_{i},v_{j}\right)$. A matrix with
positive or zero off-diagonal elements is referred to as a Metzler
matrix \cite{luenberger1979}. The Metzler matrix $L$ for the graph
$\mathcal{G}$ is defined as $L\triangleq A-D\in\mathbb{R}^{N\times N}$,
where $D\triangleq\left[d_{ij}\right]\in\mathbb{R}^{N\times N}$ is
a diagonal matrix with $d_{ii}=\sum\nolimits _{j=1}^{N}a_{ij}.$ To
facilitate the following development, a Lemma in \cite{moreau2004}
is introduced as follows.
\begin{lem}
\label{Lemma1}\cite{moreau2004} Consider a linear system $\dot{x}\left(t\right)=A\left(t\right)x\left(t\right),$
and a Lyapunov function $V\left(x\right)=\max\left\{ \begin{array}{ccc}
x_{1}, & \cdots, & x_{n}\end{array}\right\} -\min\left\{ \begin{array}{ccc}
x_{1}, & \cdots, & x_{n}\end{array}\right\} ,$ where $x\left(t\right)=\left[x_{1},\ldots,x_{n}\right]^{T}\in$$R^{n}$
is a $n$ dimensional state. If the time-varying matrix $A\left(t\right)\in R^{n\times n}$
is a piecewise continuous function of time with bounded elements,
$A\left(t\right)$ is a Metzler matrix with zero row sums, \textcolor{black}{and
the time-varying graph corresponding to $A\left(t\right)$ has a spanning
tree for }all $t\geq0$, then $\dot{V}\leq0$ for all $t\geq0$ and
consensus is achieved, i.e., $x_{1}=\cdots=x_{n}$. 
\end{lem}

\section{Problem Formulation\label{sec:problem}}

\subsection{Individual Social Behavior}

Consider a social network composed of $n$ individuals. Each individual
$i$ maintains a state $q_{i}(t)\in\mathbb{R}^{d}$ in a social network,
which can represent opinions on social events, or human emotional
states such as happiness, love, anger or fear. It is assumed that
the current state $q_{i}(t)$ of an individual $i$ can be detected
from other social neighbors such as close friends or family in the
social network. Generally, the opinions or emotional states formed
by individuals about social events are not only influenced by the
information gathered through communication with their social neighbors,
but also depend on their personal experiences. To capture the evolution
of individual social states by taking into account not only exogenous
influence (e.g., information from friends or family) but also their
own character (e.g., past experience, memory), \textcolor{black}{inspired
by the works of \cite{Ahmad2007,song2010happiness,Cucker2007,Blondel2009},}
$q_{i}(t)$ is modeled as the solution to a fractional-order dynamics
as
\begin{equation}
_{0}D_{t}^{\alpha}q_{i}(t)=u_{i}(t),\text{ }i=1,\cdots,n,\label{eq:dynamics}
\end{equation}
where $u_{i}\in\mathbb{R}^{d}$ denotes an influence (i.e., control
input) over the social state, and $_{0}D_{t}^{\alpha}q_{i}(t)$ is
the $\alpha^{th}$ derivative of $q_{i}(t)$ with $\alpha\in(0,1]$.

Note that the model in (\ref{eq:dynamics}) is a heuristic approximation
to a social response, which indicates that a person's social state
has a direct relationship with external influence integrated over
the history of a person's emotional states. On-going efforts by the
scientific community are focused on the development of clinically
derived models; yet, to date, there is no widely accepted model to
describe a person's social response in a social network.

\subsection{Social Interaction}

Let $\mathcal{G}\left(t\right)=(\mathcal{V},\mathcal{E}\left(t\right))$
denote a directed graph, where the node set $\mathcal{V}=\left\{ 1,\cdots,n\right\} $
represents individuals and the edge set $\mathcal{E}\subset\mathcal{V\times V}$
represents the interactions between individuals in a social network.
Suppose that there exist $m$ followers $i\in\mathcal{V}_{F}$, $i=1,\cdots,m$,
and $n-m$ leaders $j\in\mathcal{V}_{L}$, $j=m+1,\cdots,n$, where
$\mathcal{V}_{L}$ and $\mathcal{V}_{F}$ denote the leader and the
follower set, respectively, satisfying $\mathcal{V}_{L}\cup\mathcal{V}_{F}=\mathcal{V}$.
It is assumed that the leaders' states $q_{i}\in\mathbb{R}^{d},$
$i\in\mathcal{V}_{L},$ are desired and immutable. For each follower,
its state $q_{i}(t),$ $i\in\mathcal{V}_{F},$ evolves according to
the dynamics (\ref{eq:dynamics}) under the influence from both followers
and leaders directly or indirectly by the underlying network.

A directed edge $\left(i,j\right)\in\mathcal{E}$ in $\mathcal{G}$
represents the neighborhood of node $i$ and $j$. Each edge $\left(i,j\right)$
is associated with a time-varying weighting factor called the social
difference $S_{ij}\in\mathbb{R}^{+},$ which is defined as $S_{ij}=\left\Vert q_{i}-q_{j}\right\Vert ^{2}.$
Since individuals are assumed that they fail to incorporate the information
provided by neighbors whose states are far from their own, the designed
social difference $S_{ij}$ aims to capture the closeness of the states
between two neighboring nodes $i$ and $j.$ It is also assumed that
there exists a threshold $\mathbb{\delta}\in\mathbb{R}^{+}$, and
individuals $i$ and $j$ are able to influence each other only when
their social difference $S_{ij}\leq\mathbb{\delta}$. In other words,
an edge $\left(i,j\right)$ in graph $\mathcal{G}$ does not exist
if the social difference $S_{ij}$ is greater than the threshold $\mathbb{\delta}$.
The neighbors of individual $i$ in graph $\mathcal{G}$ are defined
as $\mathcal{N}_{i}=\left\{ j\text{ }|\text{ }S_{ij}\leq\mathbb{\delta}\right\} $,
which determines a set of individuals who can influence the social
states of individual $i.$ A directed path from node $v_{1}$ to node
$v_{k}$ is a sequence of edges $\left(v_{1},v_{2}\right),$ $\left(v_{2},v_{3}\right),\cdots,$
$\left(v_{i},v_{k}\right)$ in the directed graph. If graph $\mathcal{G}$
contains a directed spanning tree, every node has exactly one parent
node except for one node, called the root, and the root has directed
paths to every other node in graph $\mathcal{G}$.

\textbf{Assumption 1: }For each follower $i\in\mathcal{V}_{F},$ there
exists at least one leader that has a directed path to the follower
$i$ in the initial graph $\mathcal{G}\left(0\right)$.

Assumption 1 implies that there exists a directed spanning tree for
the initial graph $\mathcal{G}\left(0\right)$, where the set of leaders
acting as the roots in the directed spanning tree has an influence
directly or indirectly on all followers through a series of directed
paths in the network. Note that a connected graph (i.e., a directed
tree structure) is only assumed in the initial graph, and the controller
developed in the subsequent section will preserve the network connectivity
to ensure consistent influence between social neighbors.

\subsection{Objectives}

Let $\mathbf{q}\left(t\right),$ $\mathbf{q}^{F}\left(t\right)$,
and $\mathbf{q}^{L}\left(t\right)$ denote the stacked vector of all
states $q_{i}\left(t\right),$ $i\in\mathcal{V}$, the followers'
states $q_{i}\left(t\right),$ $i\in\mathcal{V}_{F},$ and the leaders'
states $q_{i}\left(t\right)$, $i\in\mathcal{V}_{L}$, respectively.
The convex hull spanned by the states of leaders, and all states (i.e.,
both leaders and followers), are then represented as $Co\left(\mathbf{q}^{L}\left(t\right)\right)$
and $Co\left(\mathbf{q}\left(t\right)\right),$ respectively. Since
the leaders' states are static, the convex hull $Co\left(\mathbf{q}^{L}\left(t\right)\right)$
is constant, while the convex hull $Co\left(\mathbf{q}\left(t\right)\right)$
is time varying and depends on the states of the followers. After
formulating the social network as a networked fractional-order system
described by (\ref{eq:dynamics}), the objective is to regulate the
states of followers to a desired region, which is a convex hull spanned
by all stationary leaders' states (i.e., $q_{i}(t)\rightarrow Co\left(\mathcal{V}_{L}\right)$
$\forall i\in\mathcal{V}_{F}$). To ensure that each individual is
able to be influenced by social leaders through a path of directed
edges by communication with their local neighbors only, another goal
is to preserve the network connectivity for the underlying social
network (i.e., maintain the social difference $S_{ij}\leq\mathbb{\delta}$
so that peers remain peers) when given an initially connected graph
$\mathcal{G}$. Since the systems in (\ref{eq:dynamics}) along different
dimensions are decoupled, for the simplicity of presentation, only
a scalar system ($d=1$), that is $q_{i}(t)\in\mathbb{R}$, is considered
in the following analysis. However, the results are valid for a $d$
dimensional case by the introduction of the Kronecker product.

\section{Distributed Influence Design\label{sec:controller}}

The artificial potential field based approach is one of the most widely
used methods in the control of multi-agent systems, which consists
of an attractive potential encoding the control objective and a repulsive
potential representing the motion constraints (cf. \cite{Rimon_1990}).
To apply the potential field based approach to a social network problem,
inspired by the work of \cite{Dimarogonas2010} and \cite{Kan.Dani.ea2012},
a decentralized potential function $\varphi_{i}:\mathbb{R}^{d}\rightarrow[0,1]$
$\forall i\in\mathcal{V}_{F}$ is developed to influence the followers'
states to a desired end as
\begin{equation}
\varphi_{i}=\frac{\gamma_{i}}{\left(\gamma_{i}^{k}+\beta_{i}\right)^{1/k}},\text{ }i\in\mathcal{V}_{F}\label{eq: Navigation fcn}
\end{equation}
where $k\in\mathbb{R}^{+}$ is a tuning parameter, $\gamma_{i}:\mathbb{R}^{d}\rightarrow\mathbb{R}^{+}$
is the goal function, and $\beta_{i}:\mathbb{R}^{d}\rightarrow\mathbb{R}^{+}$
is a constraint function. The goal function in (\ref{eq: Navigation fcn})
is designed as
\begin{equation}
\gamma_{i}=\sum\nolimits _{j\in\mathcal{N}_{i}}\frac{1}{2}\left\Vert q_{i}-q_{j}\right\Vert ^{2},\label{goal fcn}
\end{equation}
which aims to achieve consensus for node $i$ with its neighbors $j\in\mathcal{N}_{i}$.
To ensure consistent influence from neighbors (i.e., maintain the
social difference $S_{ij}\leq\mathbb{\delta}$), the constraint function
in (\ref{eq: Navigation fcn}) is designed as 
\begin{equation}
\beta_{i}=\frac{1}{2}\prod\nolimits _{j\in\mathcal{N}_{i}}b_{ij},\label{eq:beta_i}
\end{equation}
where $b_{ij}=\mathbb{\delta}-S_{ij}\in\mathbb{R}^{+}$. For an existing
interaction between individuals $i$ and $j,$ the potential function
$\varphi_{i}$ in (\ref{eq: Navigation fcn}) will approach its maximum
whenever the constraint function $\beta_{i}$ decreases to $0$ (i.e.,
the social difference $S_{ij}$ increases to the threshold of $\mathbb{\delta}$).

Based on the definition of the potential function in (\ref{eq: Navigation fcn}),
the distributed influence algorithm for each follower is designed
as 
\begin{equation}
u_{i}=-K_{i}\nabla_{q_{i}}\varphi_{i},\text{ }i\in\mathcal{V}_{F}\label{controller}
\end{equation}
where $K_{i}$ is a positive gain, and $\nabla_{q_{i}}\varphi_{i}$
denotes the gradient of $\varphi_{i}$ with respect to $q_{i}$. Applying
(\ref{controller}) to (\ref{eq:dynamics}), the closed-loop dynamics
of social response for all individuals in a social network can be
obtained as
\begin{equation}
\left\{ \begin{array}{cc}
_{0}D_{t}^{\alpha}q_{i}(t)=-K_{i}\nabla_{q_{i}}\varphi_{i} & \text{ }i\in\mathcal{V}_{F}\\
_{0}D_{t}^{\alpha}q_{i}(t)=0 & \text{ }i\in\mathcal{V}_{L}.
\end{array}\right.\label{cl_dynamics}
\end{equation}
Since leaders' states are stationary, the input to leaders in (\ref{cl_dynamics})
is zero, and $\nabla_{q_{i}}\varphi_{i}$ can be computed as 
\begin{equation}
\nabla_{q_{i}}\varphi_{i}=\frac{k\beta_{i}\nabla_{q_{i}}\gamma_{i}-\gamma_{i}\nabla_{q_{i}}\beta_{i}}{k(\gamma_{i}^{k}+\beta_{i})^{\frac{1}{k}+1}}.\label{eq:phi_gd}
\end{equation}
From (\ref{goal fcn}) and (\ref{eq:beta_i}), $\nabla_{q_{i}}\gamma_{i}=\sum\nolimits _{j\in\mathcal{N}_{i}}\left(q_{i}-q_{j}\right)$
and $\nabla_{q_{i}}\beta_{i}=-\sum\nolimits _{j\in\mathcal{N}_{i}}\bar{b}_{ij}\left(q_{i}-q_{j}\right)$
respectively, where $\bar{b}_{ij}\triangleq\prod\nolimits _{l\in\mathcal{N}_{i},l\neq j}b_{il}\in\mathbb{R}^{+}.$
Substituting $\nabla_{q_{i}}\gamma_{i}$ and $\nabla_{q_{i}}\beta_{i}$
into (\ref{eq:phi_gd}), $\nabla_{q_{i}}\varphi_{i}$ is rewritten
as
\begin{equation}
\nabla_{q_{i}}\varphi_{i}=\sum\nolimits _{j\in\mathcal{N}_{i}}m_{ij}\left(q_{i}-q_{j}\right),\label{eq:phi_gd2}
\end{equation}
where 
\begin{equation}
m_{ij}=\frac{k\beta_{i}+\bar{b}_{ij}\gamma_{i}}{k(\gamma_{i}^{k}+\beta_{i})^{\frac{1}{k}+1}}\label{eq:m_ij}
\end{equation}
is non-negative, based on the definition of $\gamma_{i}$, $\beta_{i}$,
$k$, and $\bar{b}_{ij}$.

\section{Convergence Analysis \label{sec:analysis}}

To show that the followers in the fractional-order network converge
to a convex hull spanned by the static leaders' states, the following
analysis is segregated into three proofs. The first proof shows that
the existing interaction between individuals is maintained by the
influence function designed in (\ref{controller}) (i.e., the social
difference $S_{ij}\leq\mathbb{\delta}$ for all time, meaning influential
peers remain influential), and thus the connectivity for the social
network is preserved. The second proof yields the asymptotic stability
for an integer-order representation of the dynamic system in (\ref{eq:dynamics}),
which is then used to establish the asymptotic convergence to the
equilibrium set of consensus states for the fractional-order system
by using a Mittag-Leffler stability analysis in the third proof.

\subsection{Maintenance of Social Influence}

If a directed graph $\mathcal{G}\left(t\right)$ does not have a directed
spanning tree, there must exist a follower to which all leaders do
not have a directed path to influence the follower's states. Hence,
the state of the follower is independent of the influence of leaders,
and thus can not converge to the stationary convex hull spanned by
leaders. To ensure the continued influence from leaders to all followers,
a directed spanning tree structure must be maintained all the time.
The following theorem shows that, given an initial graph containing
a directed spanning tree assumed in Assumption 1, the tree structure
will be preserved under the influence function in (\ref{controller})
(i.e., network connectivity is maintained and social peers do not
become isolated from the social group).
\begin{thm}
\label{Thm1}The influence function in (\ref{controller}) guarantees
a directed spanning tree structure in $\mathcal{G}$\ for all time. \end{thm}
\begin{pf}
It is assumed that the initial graph $\mathcal{G}\left(0\right)$
has a directed spanning tree in Assumption 1. If every existing edge
in $\mathcal{G}\left(0\right)$ is preserved, the tree structure will
also be preserved. Since an individual only considers a local neighbor's
opinion as reasonable when their social difference $S_{ij}\leq\mathbb{\delta}$,
peer influence is maintained when each edge $S_{ij}\leq\mathbb{\delta}$
all the time. Consider a state $q_{0}$ for individual $i$, where
the interaction between individual $i$ and neighbor $j\in\mathcal{N}_{i}$
satisfies $b_{ij}\left(q_{0},q_{j}\right)=0,$ which indicates that
their social difference is too large to influence each others' opinion,
and the associated edge is about to break. From (\ref{eq:beta_i}),
$\beta_{i}=0$ when $b_{ij}=0$, and the navigation function $\varphi_{i}$
achieves its maximum value from (\ref{eq: Navigation fcn}). Since
$\varphi_{i}$ is maximized at $q_{0},$ no open set of initial conditions
can be attracted to $q_{0}$ under the negative gradient control law
designed in (\ref{controller}). Therefore, the social bond between
individual $i$ and $j$ is maintained less than $\delta$ by (\ref{controller}),
and the associated edge is also maintained. Repeating this argument
for all pairs, every edge in $\mathcal{G}$ is maintained and the
directed spanning tree structure is preserved. \qed
\end{pf}

\subsection{Convergence Analysis}

To establish asymptotic convergence to the equilibrium points (i.e.,
the convex hull $Co\left(\mathbf{q}^{L}\left(t\right)\right)$ for
the fractional-order dynamics in (\ref{eq:dynamics}), an integer-order
system $\dot{q}_{i}(t)=u_{i}(t)$ with $\alpha=1$ in (\ref{cl_dynamics})
is considered first in the following theorem.
\begin{thm}
\label{Thm2}Consider a network composed of stationary leaders and
dynamic followers described by (\ref{cl_dynamics}). The followers
$i\in V_{F}$\ asymptotically converge to the equilibrium points
(i.e., a convex hull $Co\left(\mathbf{q}^{L}\left(t\right)\right)$
spanned by the leaders states only), if there always exists at least
one leader $j\in V_{L}$\ that has a directed path to any follower
$i$\ (i.e., a directed spanning tree is maintained). \end{thm}
\begin{pf}
This theorem is proven by using LaSalle's invariance principle and
convex properties. Let $V\left(\mathbf{q}\left(t\right)\right)$ be
the volume of the convex hull $Co\left(\mathbf{q}\left(t\right)\right)$
formed by all leaders' and followers' states. First, we show that
there exists a compact set $\Omega$ such that if $q_{i}\left(0\right)\in\Omega$
for $\forall i\in\mathcal{V}$, then $q_{i}\left(t\right)\in\Omega$
for all $t\geq0$, which implies that $\Omega$ is a positively invariant
set. Second, let $E$ be the set of all points in $\Omega$ where
$\dot{V}=0$ (i.e., the volume of $Co\left(\mathbf{q}\left(t\right)\right),$
$\mathbf{q}\left(t\right)\in E$, stays constant). It is then shown
that $M$ is the largest invariant set, where $M$ is the set of points
in the convex hull $Co\left(\mathbf{q}^{L}\left(t\right)\right)$
formed by stationary leaders only.

Substituting (\ref{eq:phi_gd2}) into (\ref{cl_dynamics}) with $\alpha=1$
yields the following closed-loop emotion dynamics
\begin{equation}
\left\{ \begin{array}{cc}
\dot{q}_{i}(t)=-\sum\limits _{j\in\mathcal{N}_{i}}K_{i}m_{ij}\left(q_{i}\left(t\right)-q_{j}\left(t\right)\right) & \text{ }i\in\mathcal{V}_{F}\\
\dot{q}_{j}(t)=0 & \text{ }j\in\mathcal{V}_{L},
\end{array}\right.\label{pf2_3}
\end{equation}
which can be rewritten in a compact form of a time-varying linear
system as
\begin{equation}
\mathbf{\dot{q}}\left(t\right)=\left[\begin{array}{c}
\mathbf{\dot{q}}^{F}\left(t\right)\\
\mathbf{\dot{q}}^{L}\left(t\right)
\end{array}\right]=\left[\begin{array}{c}
\pi\left(t\right)\\
\mathbf{0}_{(n-m)\times n}
\end{array}\right]\mathbf{q}\left(t\right),\label{pf2_4}
\end{equation}
where $\mathbf{0}_{(n-m)\times n}$ denotes the $\left(n-m\right)\times n$
matrix with all zeros, and the elements of $\pi\left(t\right)\in\mathbb{R}^{m\times n}$
are defined as
\begin{equation}
\pi_{ik}\left(t\right)=\left\{ \begin{array}{cc}
-\sum\limits _{j\in\mathcal{N}_{i}}K_{i}m_{ij} & i=k\\
K_{i}m_{ik} & k\in\mathcal{N}_{i},i\neq k\\
0, & k\notin\mathcal{N}_{i},i\neq k.
\end{array}\right.\label{pf2_5}
\end{equation}
Each follower $i\in\mathcal{V}_{F}$ in (\ref{pf2_3}) evolves according
to the dynamics:
\begin{equation}
\dot{q}_{i}(t)=-\sum\limits _{j\in\mathcal{N}_{i}}\pi_{ij}\left(t\right)\left(q_{i}\left(t\right)-q_{j}\left(t\right)\right).\label{pf2_1}
\end{equation}
To facilitate the analysis, (\ref{pf2_1}) can be written in discrete
time as 
\begin{equation}
q_{i}(t+1)=\left(1-T\sum\limits _{j\in\mathcal{N}_{i}}\pi_{ij}\right)q_{i}\left(t\right)+T\sum\limits _{j\in\mathcal{N}_{i}}\pi_{ij}q_{j}\left(t\right),\label{pf2_2}
\end{equation}
where $T$ is a sufficiently small sampling period. From (\ref{pf2_2}),
it is clear that $q_{i}(t+1)$ is a convex combination of its current
state $q_{i}\left(t\right)$ and its neighbors' states $q_{j}\left(t\right)$,
$j\in\mathcal{N}_{i}$, which implies that the follower $i$ moves
towards the convex hull spanned by itself and its neighborhood set
$\mathcal{N}_{i}$. Since the leaders' states are stationary and the
followers' states are evolving within the convex hull, $V\left(\mathbf{q}\left(t\right)\right)$
is uniformly non-increasing and thus $V\left(\mathbf{q}\left(0\right)\right)$
is the compact set $\Omega$.

The next step is to show that all followers' states will asymptotically
converge to their equilibrium points. To see that the equilibrium
points are indeed the stationary convex hull $Co\left(\mathbf{q}^{L}\left(t\right)\right)$,
let $q_{i,eq}$ be an equilibrium point for a follower $i\in\mathcal{V}_{F}$.
For an equilibrium point, it must have $\dot{q}_{i,eq}=0,$ and (\ref{pf2_1})
can be written as $-\sum_{j\in\mathcal{N}_{i}}\pi_{ij}\left(t\right)\left(q_{i,eq}\left(t\right)-q_{j,eq}\left(t\right)\right)=0$,
which yields that
\begin{equation}
q_{i,eq}=\frac{1}{-\pi_{ii}\left(t\right)}\sum_{j\in\mathcal{N}_{i}}\pi_{ij}\left(t\right)q_{j,eq}\left(t\right)\label{pf2_6}
\end{equation}
by using (\ref{pf2_1}). Due to the fact that $-\pi_{ii}\left(t\right)=\sum_{j\in\mathcal{N}_{i}}\pi_{ij}\left(t\right)\in\mathbb{R}^{+}$
from (\ref{pf2_5}), (\ref{pf2_6}) indicates that the equilibrium
point $q_{i,eq}$ lies in a convex hull spanned by its neighbors'
states (i.e., leaders and/or followers). Since every follower ends
up in a convex hull spanned by its neighbors' states and the leaders'
states are static, every follower will end up in a convex hull spanned
by the leaders states only (i.e., $Co\left(\mathbf{q}^{L}\left(t\right)\right)$).
Using the fact that $m_{ij}$ is non-negative from (\ref{eq:m_ij})
and $K_{i}$ is a positive constant gain in (\ref{controller}), $\left[\begin{array}{c}
\pi\left(t\right)\\
\mathbf{0}_{(n-m)\times n}
\end{array}\right]$ in (\ref{pf2_4}) is a Metzler matrix with zero row sums. According
to Lemma \ref{Lemma1} and following a similar procedure as in \cite{moreau2004},
the convex hull $Co\left(\mathbf{q}\left(t\right)\right)$ is shrinking
(i.e., $\dot{V}\left(\mathbf{q}\left(t\right)\right)<0$), since the
difference of the extremes $\max\left\{ \begin{array}{ccc}
x_{1}, & \cdots, & x_{n}\end{array}\right\} $ and $\min\left\{ \begin{array}{ccc}
x_{1}, & \cdots, & x_{n}\end{array}\right\} $ is decreasing. If all followers states are initially within the convex
hull $Co\left(\mathbf{q}^{L}\left(t\right)\right),$ the states will
always stay within $Co\left(\mathbf{q}^{L}\left(t\right)\right)$
(i.e., $\dot{V}=0$).

A proof by contradiction can now be used to show that $M$ (i.e.,
$Co\left(\mathbf{q}^{L}\left(t\right)\right)$) is the largest invariant
set. Let $M^{\prime}\supset M$ be a larger invariant set in $E$.
Suppose that there is a follower whose state $q_{i}\left(0\right)\notin M$,
and $q_{i}\left(0\right)$ is on the boundary of $M^{\prime}.$ Since
$M^{\prime}\subset E,$ the volume of the set $M^{\prime}$ stays
constant. The only way for the volume of $M^{\prime}$ to stay constant
is that $q_{i}\left(0\right)=q_{i}\left(t\right)$ for all $t\geq0.$
However, for this to happen, we must have $\pi_{ij}\left(t\right)=0$
for $\forall j\in\mathcal{N}_{i}$ from (\ref{pf2_1})$,$ which indicates
that the follower $i$ is isolated from the group. This isolation
is a contradiction with network connectivity. Hence, $M$ is the largest
invariant set. The followers asymptotically converge to the largest
invariant set $M$ (i.e., the equilibrium points $Co\left(\mathbf{q}^{L}\left(t\right)\right)$)
by using LaSalle's invariance principle in \cite{Khalil2002}. \qed
\end{pf}
Since asymptotic stability for the integer-order system (\ref{pf2_3})
is established in Theorem \ref{Thm2}, a similar proof procedure in
our recent work \cite{Kan.Shea.ea2012} can be followed to prove asymptotic
stability for the fractional order system in (\ref{cl_dynamics})
by using Mittag-Leffler stability analysis and a Converse Lyapunov
Theorem.
\begin{thm}
\label{Thm3} The follower $i\in V_{F}$, with closed-loop fractional-order
dynamics in (\ref{cl_dynamics}) with $\alpha\in(0,1)$, asymptotically
converges to the convex hull spanned by stationary leaders if at least
one leader $j\in V_{L}$ has a directed path to the follower $i$. \end{thm}
\begin{pf}
Let $x_{i}\left(t\right)\triangleq q_{i}\left(t\right)-q_{i,eq},$
and $\mathbf{x}^{F}\left(t\right)\triangleq\mathbf{q}^{F}\left(t\right)-\mathbf{q}_{eq}^{F}$,
where $\mathbf{q}_{eq}^{F}$ denotes the stacked vector of $q_{i,eq}$.
Since the leaders' states are constant, the closed-loop fractional-order
dynamics in (\ref{cl_dynamics}) can be written in a compact form
as 
\begin{equation}
_{0}D_{t}^{\alpha}\mathbf{x}^{F}\left(t\right)=g(\mathbf{x}^{F}),\label{pf3_0}
\end{equation}
for all followers where $g(\mathbf{x}^{F})$ is a function of follower
states. Since stability of a fractional-order system is defined by
Definition \ref{def1}, and Mittag-Leffler stability implies asymptotic
convergence as discussed in \cite{li2009mittag}, the following development
aims to show that (\ref{pf3_0}) is Mittag-Leffler stable.

Since asymptotic stability is established in Theorem \ref{Thm2} for
the integer-order system of (\ref{pf2_2}), a Converse Lyapunov Theorem,
(i.e., Theorem 4.16 in \cite{Khalil2002}) is invoked to establish
that there exists a function $V\left(t,\mathbf{x}^{F}\right):\left(0,\infty\right]\times\mathbb{R}^{m}\mathbb{\rightarrow R}$
and class $\mathcal{K}$ functions $\sigma_{i}$ ($i=1,2,3$) such
that 
\begin{equation}
\sigma_{1}\left(\left\Vert \mathbf{x}^{F}\right\Vert \right)\leq V\left(t,\mathbf{x}^{F}\right)\leq\sigma_{2}\left(\left\Vert \mathbf{x}^{F}\right\Vert \right),\label{pf3_1}
\end{equation}
\begin{equation}
\dot{V}\leq-\sigma_{3}\left(\left\Vert \mathbf{x}^{F}\right\Vert \right).\label{pf3_2}
\end{equation}
Let $\beta=1-\alpha\in(0,1)$. From Theorem 8 in \cite{li2009mittag}
and (\ref{pf3_2}), the fractional derivative of $V$ is computed
as
\begin{eqnarray}
_{0}D_{t}^{\beta}V\left(t,\mathbf{x}^{F}\right) & = & \text{ }_{0}D_{t}^{1-\alpha}V\left(t,\mathbf{x}^{F}\right)=\text{ }_{0}D_{t}^{-\alpha}\dot{V}\\
 & \leq & \text{ }-_{0}D_{t}^{-\alpha}\left(\sigma_{3}\left(\left\Vert \mathbf{x}^{F}\right\Vert \right)\right).\label{pf3_4}
\end{eqnarray}
From the definition of the fractional integral $_{0}D_{t}^{-\alpha}f\left(t\right)=\frac{1}{\Gamma\left(\alpha\right)}\int_{0}^{t}\frac{f\left(\tau\right)}{\left(t-\tau\right)^{1-\alpha}}d\tau$,
where $\Gamma\left(\cdot\right)$ denotes the Gamma function \cite{monje2010},
it is known that $_{0}D_{t}^{-\alpha}\left(\cdot\right)$ is a class
$\mathcal{K}$ function, since $_{0}D_{t}^{-\alpha}\left(0\right)=0$
and $_{0}D_{t}^{-\alpha}\left(\cdot\right)$ is strictly increasing
on the domain $\left[0,\infty\right)$. Using the fact that $a_{1}\circ a_{2}$
also belongs to class $\mathcal{K}$, where $a_{1}$ and $a_{2}$
are class $\mathcal{K}$ functions, the term $_{0}D_{t}^{-\alpha}\left(\sigma_{3}\left(\left\Vert \mathbf{x}^{F}\right\Vert \right)\right)$
in (\ref{pf3_4}) is a class $\mathcal{K}$ function, since both $_{0}D_{t}^{-\alpha}\left(\cdot\right)$
and $\sigma_{3}\left(\cdot\right)$ are class $\mathcal{K}$ functions.
Thus, the inequality in (\ref{pf3_4}) can be written as
\begin{equation}
_{0}D_{t}^{\beta}V\left(t,\mathbf{x}^{F}\right)\leq-\sigma_{4}\left(\left\Vert \mathbf{x}^{F}\right\Vert \right),\label{pf3_5}
\end{equation}
where $\sigma_{4}\left(\left\Vert \mathbf{x}^{F}\right\Vert \right)\triangleq$
$_{0}D_{t}^{-\alpha}\left(\sigma_{3}\left(\left\Vert \mathbf{x}^{F}\right\Vert \right)\right)$
is a class $\mathcal{K}$ function. Applying Lemma \ref{lemma2} to
(\ref{pf3_1}) and (\ref{pf3_5}), Mittag-Leffler stability of (\ref{pf3_0})
with $\alpha\in(0,1)$ can be obtained, which implies that the equilibrium
points $\mathbf{q}_{eq}^{F}$ for the followers in the closed-loop
fractional-order system in (\ref{pf3_0}) are asymptotically stable.
\qed
\end{pf}

\section{Simulations}

\begin{figure}
\centering{}\includegraphics[scale=0.8]{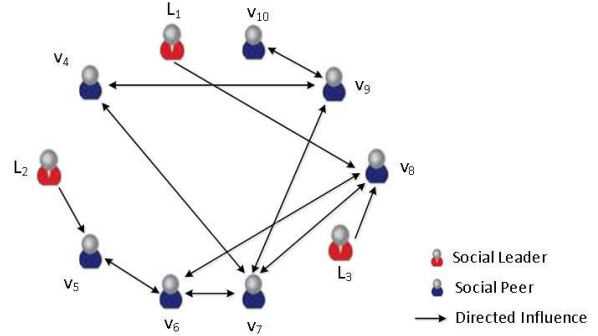}

\caption{The Zachary's karate club network is modeled by a directed graph. }

\label{fig:karate}
\end{figure}

\begin{figure}
\centering{}\includegraphics[scale=0.5]{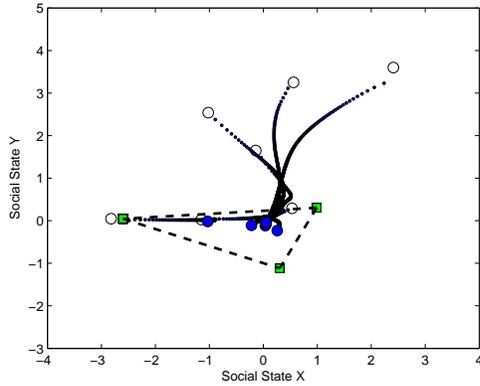}

\caption{Evolution of social states. The followers' states converge to the
convex hull formed by social leaders, where the the leaders' social
states are denoted as squares, and the followers' initial and final
social states are denoted by circles and dots, respectively.}

\label{fig:trajectory}
\end{figure}

To illustrate the proposed control algorithms, simulations are performed
on a karate club network described in \cite{zachary1977karate}. The
karate club network considered in this section consists of 3 social
leaders and 7 followers, and is represented as a directed graph in
Fig. \ref{fig:karate}. The solid arrow connecting two individuals
in Fig. \ref{fig:karate} indicates an established social bond (e.g.,
friendship) and the directed influence between individuals. Note that
the leaders can not be influenced, while the followers can be influenced
by social peers as well as social leaders. Based on the topology described
in Fig. \ref{fig:karate}, each individual is randomly assigned a
social state (e.g., an opinion on an event or an emotional state such
as happiness, fear and anger). Without loss of generality, we assume
that the social states of individuals are two dimensional (i.e., $q_{i}(t)\in\mathbb{R}^{2}$).
The control law in (\ref{controller}) yields the simulation results
shown in Fig. \ref{fig:trajectory}, which illustrates that the followers'
states converge to the convex hull formed by the social leaders%
\footnote{\textcolor{black}{An infinite state model is developed in \cite{Trigeassou2011}
to represent a Fractional-order Differential Equation (FDE), as a
means to solve the initial condition challenges associated with FDEs.
As indicated in \cite{Trigeassou2011}, the internal state $z\left(t\right)$
in the infinite state model contains historical information of the
fractional-order system. To address the initialization problem of
a FDE, the observer based model in \cite{Trigeassou2011} can be used
to estimate $z\left(t_{0}\right)$ by using past information from
$t<t_{0}$. Since the current work focuses on showing how the individual\textquoteright{}s
behavior can be influenced under the designed influence function,
for simplicity in the simulation, $z\left(t_{0}\right)$ is assumed
known, which captures the individual's historical experience. Given
that the fractional-order dynamics in (\ref{eq:dynamics}) is written
as $q_{i}\left(t\right)=I_{n}\left(u_{i}\right)$ where $I_{n}\left(u_{i}\right)$
is the $n$th fractional integral of control input $u_{i}$, the trajectory
of $q_{i}\left(t\right)$ is simulated by following the infinite state
approach in \cite{Trigeassou2011}. An alternate initialization approach
is to account for the initialization function (cf. \cite{Lorenzo2008,Sabatier2013}).}%
}.

\section{Conclusion\label{sec:conclusion}}

By modeling the group social response as a networked fractional-order
system, a decentralized potential field-based influence algorithm
is developed in this work to ensure that all individuals' states achieve
consensus asymptotically to a desired convex hull spanned by the stationary
leaders' states, while maintaining consistent influence between individuals
(i.e., network connectivity).\textcolor{black}{{} This work considers
individuals whose social response is modeled by a FOS with $\alpha\in(0,1]$.
Since some individuals may respond with a more complex dynamic (e.g.,
$\alpha\in(1,2]$), future efforts will focus on generalizing the
development to include networks with heterogeneous members with higher
order dynamic response. }Future effort will also consider different
influence capabilities between individuals. For instance, a person
tends to have a larger tolerance for a difference of opinions for
a certain social event in a close friend than a loose acquaintance,
and thus, can be more easily influenced by the close friend.

\bibliographystyle{IEEEtran}
\bibliography{master,ncr,\string"E:/nonliner control/my paper/bibtex/bib/ncrbibs/master\string",\string"E:/nonliner control/my paper/bibtex/bib/ncrbibs/ncr\string",\string"C:/ZHEN KAN/NCR/My Paper/bibtex/bib/ncrbibs/master\string",\string"C:/ZHEN KAN/NCR/My Paper/bibtex/bib/ncrbibs/ncr\string"}

\end{document}